\documentclass{emulateapj}

\newcommand{\Msol}{\ensuremath{\mathrm{M_{\odot}}}}

\newcommand{\rt}{\ensuremath{R_{\mathrm{200}}}}

\newcommand{\Zsol}{\ensuremath{\mathrm{Z_{\odot}}}}
\newcommand{\fgas}{\ensuremath{f_{\mathrm{gas}}}}

\newcommand{\OM}{\ensuremath{\Omega_{\mathrm{M}}}}

\newcommand{\egc}{{\it e.g.}}  
\newcommand{\etal}{{\it et al.\thinspace}}

\newcommand{\Chandra}{\emph{Chandra}}

\newcommand{\ROSAT}{\emph{ROSAT}}
\newcommand{\XMM}{\emph{XMM-Newton}}

\newcommand{\BeppoSAX}{\emph{BeppoSAX}}
\newcommand{\MEKAL}{\textsc{MeKaL}}

\newcommand{\chisq}{\ensuremath{\chi^2}}

\newcommand{\gta}{\,\rlap{\raise 0.4ex\hbox{$>$}}{\lower 0.6ex\hbox{$\sim$}}\,}  
\newcommand{\lta}{\,\rlap{\raise 0.4ex\hbox{$<$}}{\lower 0.6ex\hbox{$\sim$}}\,}  

\newcommand{\nm}{\mbox{\ensuremath{\mathrm{~\nm}}}}

\newcommand{\cm}{\mbox{\ensuremath{\mathrm{~cm}}}}

\newcommand{\km}{\mbox{\ensuremath{\mathrm{~km}}}}

\newcommand{\kpc}{\mbox{\ensuremath{\mathrm{~kpc}}}}
\newcommand{\Mpc}{\mbox{\ensuremath{\mathrm{~Mpc}}}}
\newcommand{\s}{\mbox{\ensuremath{\mathrm{~s}}}}
\newcommand{\ks}{\mbox{\ensuremath{\mathrm{~ks}}}}

\newcommand{\keV}{\mbox{\ensuremath{\mathrm{~keV}}}}
\newcommand{\erg}{\mbox{\ensuremath{\mathrm{~erg}}}}

\newcommand{\arcm}{\ensuremath{\mathrm{^\prime}}}
\newcommand{\arcs}{\arcm\hskip -0.1em\arcm}

\newcommand{\pcc}{\ensuremath{\mathrm{\cm^{-3}}}}

\newcommand{\pcmsq}{\mbox{\ensuremath{\mathrm{~cm^{-2}}}}}
\newcommand{\pMpc}{\ensuremath{\mathrm{\Mpc^{-1}}}}
\newcommand{\ps}{\ensuremath{\mathrm{\s^{-1}}}}

\newcommand{\ergps}{\ensuremath{\mathrm{\erg \ps}}}
\newcommand{\flux}{\ensuremath{\mathrm{\erg \ps \pcmsq}}}

\newcommand{\kmps}{\ensuremath{\mathrm{\km \ps}}}
\newcommand{\kmpspMpc}{\ensuremath{\mathrm{\km \ps \pMpc\,}}}

\newcommand{\LCDM}{$\Lambda$CDM~}

\newcommand{\jjj}{ClJ0152.7$-$1357}

\begin{document}

\title{\XMM\ observes ClJ0152.7$-$1357: A massive galaxy cluster forming at merger crossroads at $z=0.83$.}

\author{B. J. Maughan\altaffilmark{1}}
\affil{Harvard-Smithsonian Center for Astrophysics, 60 Garden St, Cambridge, MA 02140, USA.}
\altaffiltext{1}{\Chandra\ fellow}
\email{bmaughan@cfa.harvard.edu}
\author{S. C. Ellis}
\affil{Anglo-Australian Observatory, PO Box 296, Epping, NSW 2121, Australia.}
\author{L. R. Jones}
\affil{School of Physics and Astronomy, The University of Birmingham,  Edgbaston, Birmingham B15 2TT, UK.}
\author{K. O. Mason}
\affil{Mullard Space Science Laboratory, University College London,
Holmbury St. Mary, Dorking, Surrey, RH5 6NT, UK.}
\author{F. A. C\'ordova}
\affil{University of California, Riverside, CA 92521, USA}
\author{W. Priedhorsky}
\affil{Los Alamos National Laboratory, MS B241, Los Alamos, NM 87545, USA}

\shorttitle{ClJ0152.7$-$1357: A massive cluster forming at merger crossroads.}
\shortauthors{B. J. Maughan \etal}


\begin{abstract}
We present an analysis of a $50\ks$ \XMM\ observation of the
merging galaxy cluster \jjj\ at $z=0.83$. In addition to the two main
subclusters and an infalling group detected in an earlier \Chandra\
observation of the system, \XMM\ detects another group of galaxies possibly
associated with the cluster. This group may be connected to the northern
subcluster by a filament of cool ($1.4^{+0.3}_{-0.1}\keV$) X-ray
emitting gas, and lies outside the estimated virial radius of the northern
subcluster. The X-ray morphology agrees well with the projected galaxy
distribution in new K-band imaging data
presented herein. We use detailed spectral and imaging analysis of the
X-ray data to probe the dynamics of the system and find evidence that
another subcluster or group has recently passed through the northern
subcluster. \jjj\ is an extremely dynamically active system with
mergers at different stages occurring along two perpendicular merger
axes.
\end{abstract}


\section{Introduction} \label{sect:intro}
Clusters of galaxies are believed to form hierarchically via the collapse
and merger of smaller structures. Massive clusters form most recently in
this scenario, and indeed, X-ray substructure which is indicative of
merger and formation activity is found to be more prevalent in clusters at
high redshifts \citep{jel05}. Observations of merging systems at high
redshift thus enable the study of the process of cluster formation. X-ray
observations provide a useful tool in this study, enabling measurements of
the properties of the dominant baryonic component of clusters and detecting
merger-related features like shocks \citep[\egc][]{mar02} and cold fronts
\citep[\egc][]{mar00c,vik01} in the X-ray emitting gas.

In cosmological simulations, the most massive clusters form at the
intersections of the filaments of structure \citep[\egc][]{jen98}. 
While filamentary structures are observed in the large scale
distributions of galaxies \citep[\egc][]{col01}, they are rarely observed in
X-rays \citep{sch00,dur03,ebe05a} because of their low gas
densities. Cluster mergers have also been detected in X-rays at the
meeting points of apparent filaments, for example in Abell 85
\citep{dur03,dur05}, Abell 521 \citep{arn00,fer05} and Coma
\citep[\egc][]{neu03}. \jjj\ is an example of a massive, high-redshift
merging system located in a network of large-scale structures
\citep{kod05}, and so presents a valuable opportunity to observe the
hierarchical formation of structure.

\jjj\ was discovered in the Wide Angle \ROSAT\ Pointed Survey
\citep[WARPS:][]{sch97,ebe00a}. It was also discovered independently in the
RDCS \citep{ros98} and SHARC \citep{rom00} surveys and was
spectroscopically confirmed at a redshift of $z=0.83$
\citet{ebe00a,del00}. Analysis of the discovery \ROSAT\ data found the system to
be highly luminous with considerable X-ray substructure consisting of two
probable subclusters \citep{ebe00a}. The cluster has also been the subject
of  \BeppoSAX\ observations \citep{del00} and Sunyaev-Zel'dovich effect
imaging \citep{joy01}. A \Chandra\ ACIS-I observation enabled the two
subclusters to be fully resolved and detected a galaxy group near the
cluster to the east \citep{mau03a}. The \Chandra\ data were used to measure
the temperatures of the two subclusters separately for the first time,
confirming that they are both hot ($\sim5.5\keV$) and massive. 

More recently, \jjj\ has been the subject of detailed studies at optical
wavelengths. \citet{dem05} confirmed 102 cluster member galaxies with VLT
spectroscopy and found that the eastern group was at the same redshift as
the cluster. \citet{gir05} used this redshift information to perform a
detailed dynamical analysis of the system. The distribution of galaxies
around this system was probed out to scales of $\sim5\Mpc$ using
photometric redshifts obtained with Subaru by \citet{kod05}. In addition,
this remarkable system has been the subject of weak lensing analyses, both
ground based \citep{huo04} and with the HST \citep{jee05}. Broadly
speaking, the distributions of X-ray emitting gas, cluster galaxies, and
dark matter are found to be similar.

In this paper we present the results of an analysis of an \XMM\ observation
of \jjj\ along with new K-band imaging. The large collecting area of \XMM\
enables the most detailed X-ray study yet of this system. A \LCDM cosmology
of $H_0=70\kmpspMpc$, and $\OM=0.3$ ($\Omega_\Lambda=0.7$) is adopted
throughout, and all errors are quoted at the $68\%$ level. At the cluster
redshift of $0.83$, $1\arcs$ corresponds to $7.6\kpc$ in our chosen
cosmology.

\section{Data Reduction} \label{sect.data}
\subsection{\XMM\ data}
\begin{figure}
\begin{center}
\plotone{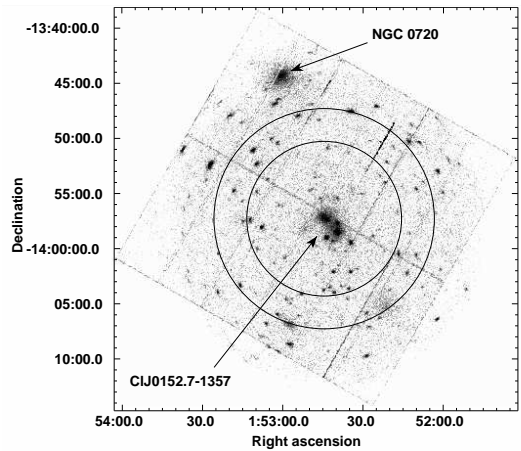}
\caption{\label{fig.fov}Binned X-ray image of the field of view. Labels
indicate the target cluster and the galaxy NGC 0720 which was observed by
chance at the edge of the field. The annulus indicates the region used for
the measurement of the background emission.} 
\end{center}
\end{figure}

\jjj\ was observed by \XMM\ for $50\ks$ on 2002 December 24 (ObsID
0109540101). The data were reduced and analysed using the \XMM\ Science
Analysis Software (SAS) version 6.1 with the latest calibration products
available in April 2005. A binned image of the field observed by \XMM\ EPIC is
shown in Fig. \ref{fig.fov}. The target cluster and the galaxy NGC 0720
(detected by chance at the edge of the field) are labelled.

Periods of background flaring were detected and removed by applying an
iterative sigma-clipping algorithm to lightcurves of the data from the PN
and two MOS detectors. This was first done for lightcurves of the whole
field in the $10-15\keV$ band with time bins of $50\s$ and a $3\sigma$
threshold.  The observation was virtually unaffected by background flares,
and after cleaning $48\ks$ ($53\ks$) of useful PN (MOS) data remained. The
results of this cleaning were checked by applying the same procedure to
lightcurves of source-free regions in the $2-15\keV$ band with $200\s$ bins
and a $2\sigma$ threshold. The results of this stricter broad-band cleaning
were completely consistent with those of the high-energy band cleaning, and
the good time intervals defined in the $10-15\keV$ band were applied to the
data used in all further analysis.

In the spectral analysis of extended sources, the precise modeling or
subtraction of the background emission is crucial, particularly in regions
of low surface brightness. The compact angular size of \jjj\ relative to
the \XMM\ field of view means that the background emission local to the
source can be used obtain background spectra. The annulus used for
measuring the local background emission is marked in Fig. \ref{fig.fov},
and has inner and outer radii of $7\arcmin$ ($3.2\Mpc$) and $10\arcmin$
($4.6\Mpc$) respectively. Note that all sources within this region were
excluded in our analysis, but these are not marked on the Figure in order
to preserve clarity. As this background region is further from the optical
axis than the source, the emission in the region will be vignetted (the
mean effective area in the background region is $\sim60\%$ of that at the
optical axis). The effects of vignetting were accounted for in the imaging
analysis with the use of exposure maps, and in all spectral analysis by
using the SAS task {\it evigweight} along with on-axis effective area
files.

\begin{figure}
\begin{center}
\includegraphics[angle=-90,width=8cm]{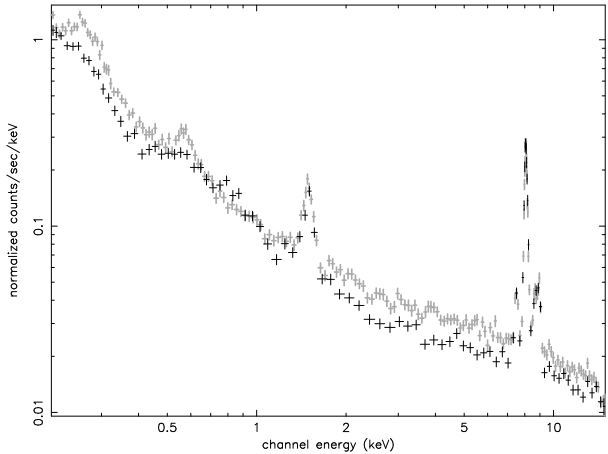}
\caption{\label{fig.bgspec}\XMM\ PN background spectra obtained from the
background annulus region in the source field (black) and blank sky dataset
(grey). The blank sky spectrum was normalised by the flux of events
detected outside the field of view.}
\end{center}
\end{figure}

All spectral analysis was also performed using the blank sky data sets of
\citet{rea03} with the ``double subtraction'' method of \citet{arn02b}. In
this method background spectra are extracted from the same detector regions
as the source spectra, but from the blank sky datasets. A background region
of the source dataset (in this case the annulus marked in
Fig. \ref{fig.fov}) is used to derive a residual spectrum to account for
any differences (predominantly at low energies due to different levels of
soft Galactic background) between the source and blank sky fields. The
final background spectrum consists of the blank sky spectrum plus the
residual spectrum (scaled for any differences in extraction area). The
exposure times of the blank sky observations were normalised so that the
count rates of events detected outside the telescopes' field of view in the
target and blank sky datasets matched. Fig. \ref{fig.bgspec} shows the PN
background spectra extracted from the background annulus region in the
source and blank sky datasets. The spectra agree well in the fluorescent
lines at $1.5\keV$ and $8\keV$ but the normalisation of the continuum in
the blank sky spectrum is too high. This indicates that the level of the
particle-induced background (which is responsible for the out of field of
view counts and dominates the fluorescent line flux) is higher in the
source dataset than the blank sky observations. Alternatively if the blank
sky spectra were normalised to match the continuum level, the normalisation
of the fluorescent lines would be incorrect. Both of these differences
would be compensated for to some extent by the residual spectrum that is
added to the blank sky spectrum.

Despite these differences, almost all of the temperatures measured with the
double subtraction blank sky background method were consistent with those
measured by adopting a local background alone (see \textsection
\ref{sect.filament} for the exception). This is most likely due to the
correcting effect of the residual spectrum, and the fact that the slope of
the blank sky background spectrum continuum is the same as the local
background. It should be noted, however, that the statistical uncertainties
on the spectral properties measured in the low surface brightness regions
of this data where the background contribution is important are large. Thus
the effects of the systematic differences between the local and blank sky
backgrounds cannot be sensitively tested. With those caveats, we conclude
that a simple local background, extracted from the source dataset, and
vignetting-corrected with {\it evigweight} is the more reliable method for
this particular dataset, and this is used for all spectral results
presented here.

\subsection{Near-infrared data}
K band observations of ClJ0152.7-1357 were made with IRIS2 (the
near-infrared imager and spectrograph) on the 3.9m Anglo-Australian
Telescope, Siding Spring, Australia, on 2003 September 5, and 2004
January 10.  The conditions on both occasions were clear, and the
seeing was $\approx 1.7\arcs$  and $1.3\arcs$ in September and January
respectively.

The images were dark-subtracted, flat-fielded and mosaiced using
standard techniques with the {\sc iraf} software package.  Flatfields
were created by median combining the jittered object frames.  The total
exposure time of all observations was 205 minutes.

\section{Imaging analysis}
A spectrally-weighted exposure map was produced for each EPIC camera, to
account for the energy dependence of the telescope vignetting
function. Exposure maps were produced in narrow energy bands, within which
the vignetting function varied little, and these were weighted according to
the relative contribution of each band to an assumed spectral model. An
absorbed \MEKAL\ model with $kT=5.5\keV$, appropriate for ClJ0152-7$-$1357
\citep[][and \textsection \ref{sect.specmap}]{mau03a}, was used. All of the
X-ray imaging analysis was performed in the $0.3-5\keV$ energy band.

\begin{figure*}
\begin{center}
\plotone{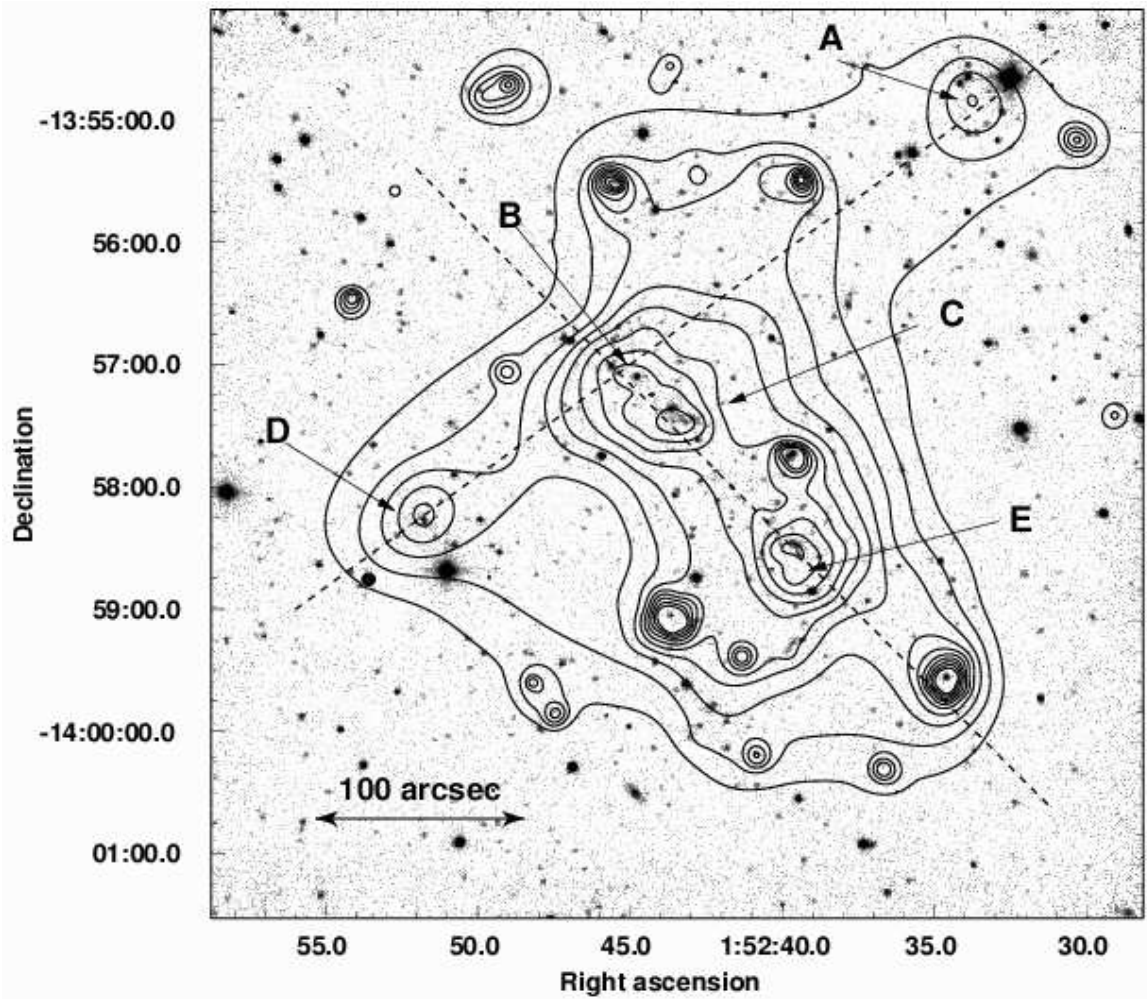}
\caption{\label{fig:overlay}Adaptively smoothed contours of the X-ray
emission detected from ClJ0152-7$-$1357 with \XMM\ are overlaid on the K
band IRIS2 image. The dashed lines indicate the two apparent merger axes of
the system and the labels are described in the text and summarised in Table
\ref{tab.labels}.} 
\end{center}
\end{figure*}

A mosaiced image of the emission detected by the three cameras was then
produced, and exposure corrected. This image was adaptively smoothed to
show real features detected at the $3\sigma$ level with the {\it asmooth}
algorithm of \citet{ebe05b}, and logarithmically spaced contours of this
smoothed emission are shown overlaid on a NIR image in
Fig. \ref{fig:overlay}. All of the sources in Fig. \ref{fig:overlay} except
for those labelled are point sources, with the possible exception of the
north eastern most source at $\alpha[2000.0]=01^{\rm h}52^{\rm m}45.64^{\rm
s}$ $\delta[2000.0]=-13^{\circ}55\arcm 29.0\arcs$. Based on the redshifts
measured by \citet{dem05}, the brightest X-ray source near the subclusters
($\alpha[2000.0]=01^{\rm h}52^{\rm m}43.62^{\rm s}$
$\delta[2000.0]=-13^{\circ}59\arcm 1.0\arcs$) appears associated with a
cluster member galaxy, and the point source at $\alpha[2000.0]=01^{\rm
h}52^{\rm m}34.62^{\rm s}$ $\delta[2000.0]=-13^{\circ}59\arcm
30.2\arcs$. Within the size of the \XMM\ point spread
function (PSF), there are two cluster member
galaxies and one foreground galaxy coincident with the point source at
$\alpha[2000.0]=01^{\rm h}52^{\rm m}39.75^{\rm s}$
$\delta[2000.0]=-13^{\circ}57\arcm 42.8\arcs$.

The complex morphology of the extended emission in \jjj\ is immediately
apparent in Fig. \ref{fig:overlay}. The north and south subclusters
(labelled C and E in Fig. \ref{fig:overlay}) and the group to the east (D)
were all detected in the \Chandra\ observation of this system
\citep{mau03a}. In addition, \XMM\ also detects an envelope of low surface
brightness emission surrounding the subclusters and eastern group, which
extends in a apparent filament to a second group (A) to the north-west. The
redshift of this group is unknown. However, while the associated galaxies
were not included in the VLT observations of \citet{dem05}, the group's
position appears to coincide with a clump of galaxies of the same
photometric redshifts as \jjj\ in \citet{kod05}. We address the question of
whether this north-west group is associated with the main cluster in more
detail later. The morphology of the system suggests intersecting north-east
to south-west and south-east to north-west merger axes, which are indicated
in Fig. \ref{fig:overlay}.
 
Fig. \ref{fig:overlay} also shows that the X-ray emission from the north
subcluster is extended in the northern direction, with a possible second
X-ray peak (labelled B). This extension is in the direction of an
overdensity of galaxies slightly further north, which also appears as a
peak in the weak lensing mass map of \citet{jee05}.

For convenience, the labelled regions in Fig. \ref{fig:overlay} are
summarised in Table \ref{tab.labels}.
\begin{deluxetable*}{cccc}
\tablecaption{\label{tab.labels}Summary of the structures labelled in Fig. \ref{fig:overlay}}
\tablehead{
\colhead{Region} & \colhead{$\alpha[2000.0]$} & \colhead{$\delta[2000.0]$} & \colhead{Description}}
\startdata
A & $01^{\rm h}52^{\rm m}34.18^{\rm s}$ & $-13^{\circ}54\arcm 42.3\arcs$ & North west group of unknown redshift \\
B & $01^{\rm h}52^{\rm m}45.15^{\rm s}$ & $-13^{\circ}57\arcm 02.4\arcs$ & Northern extension of north subcluster \\
C & $01^{\rm h}52^{\rm m}43.76^{\rm s}$ & $-13^{\circ}57\arcm 24.4\arcs$ & North subcluster \\
D & $01^{\rm h}52^{\rm m}51.99^{\rm s}$ & $-13^{\circ}58\arcm 13.2\arcs$ & Galaxy group at cluster redshift \\
E & $01^{\rm h}52^{\rm m}39.83^{\rm s}$ & $-13^{\circ}58\arcm 26.7\arcs$ & Southern subcluster
\enddata
\end{deluxetable*}

\begin{figure*}
\begin{center}
\plottwo{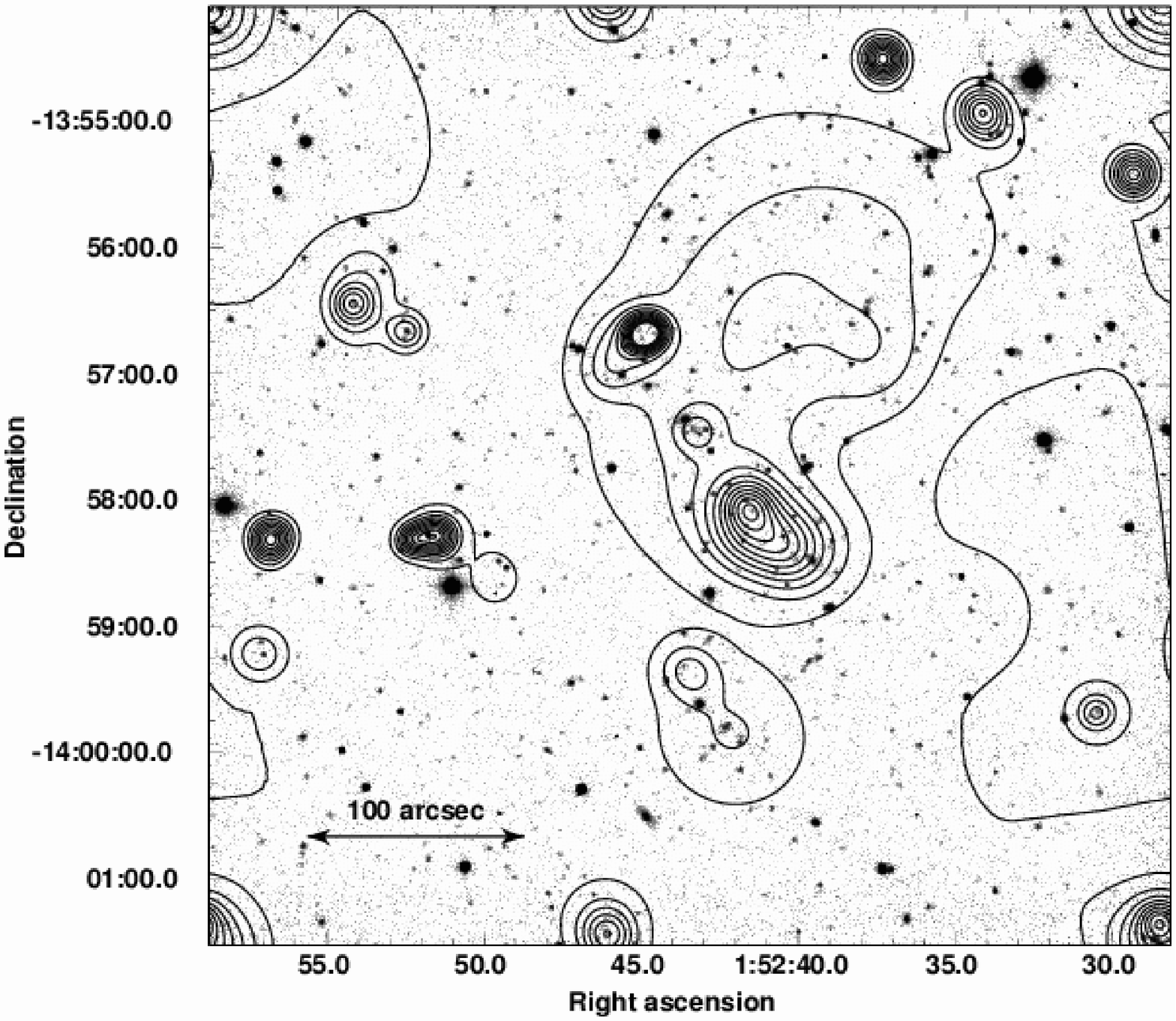}{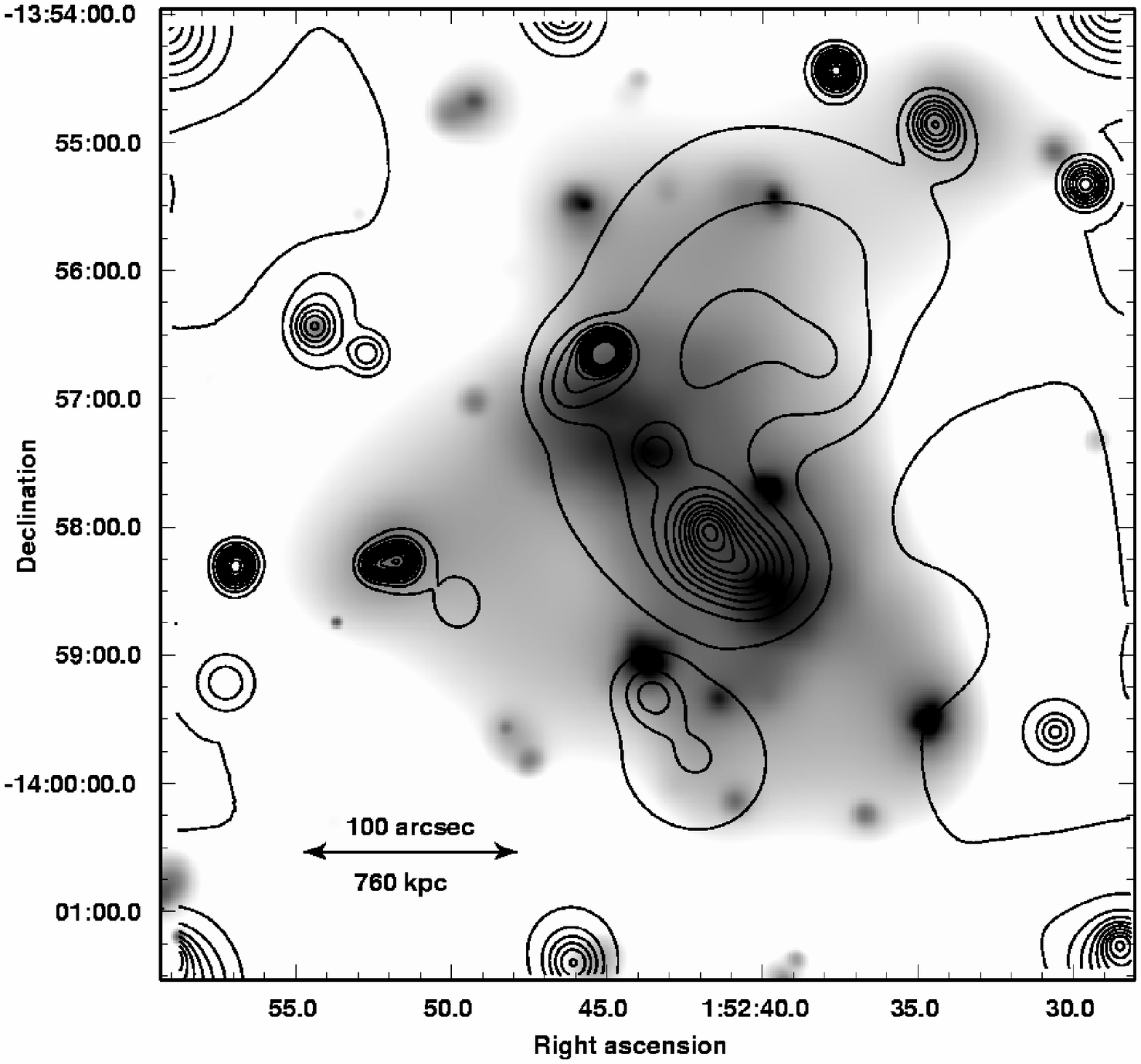}
\caption{\label{fig.galcont}{\it Left:} Contours of constant galaxy number density
derived from the AAT K-band images are overlaid on the same NIR image. The
contours have been adaptively smoothed at 90\% significance. {\it Right:}
The same contours are overlaid on the smoothed X-ray image from which the
X-ray contours in Fig. \ref{fig:overlay} were taken.}
\end{center}
\end{figure*}

The distribution of NIR light in \jjj\ was also investigated with the K
band imaging data. The NIR luminosity of a galaxy is well correlated with
its dynamical mass \citep{gav96}, making the K band an excellent choice
with which to trace out the distribution of galaxies within the
cluster. The image was divided into a grid of 1000$\times$1000 bins, and
the number of galaxies detected at a significance $\ge 3 \sigma$ in each
bin was counted. Objects which were brighter than the brightest known
members were excluded.  Due to the poor seeing, the galaxy density may be
underestimated in the highest density regions. However, all of the known
members from \citet{dem05} were included by manually deblending some of the
galaxies. An image of this map was created and adaptively smoothed at the
90\% level. Contours of galaxy number density are shown overlaid on a K
band image and a smoothed X-ray image in Fig. \ref{fig.galcont}. There is a
clear similarity in the galaxy and X-ray distribution; both the subclusters
and groups are apparent. An overdensity of galaxies to the north of the
X-ray emission in region B is also clear.

The comparison of Figs. \ref{fig:overlay} and \ref{fig.galcont} shows a
clear offset between the peak of the galaxy light and X-ray emission in the
southern subcluster. This was first noticed in the \Chandra\ observation of
\jjj\ by \citet{mau03a} who suggested that the offset was due to the
collisionless galaxies (and presumably dark matter) moving ahead of the
X-ray gas which is slowed by ram pressure due to the merger. \citet{jee05}
found similar offsets between the mass peaks of both subclusters in their
weak lensing analysis and the X-ray centroids, which is consistent with
this explanation. We note, however that the X-ray peak of the northern
subcluster is coincident with that of the galaxy distribution, but the
northern extension of that subcluster (likely due to another merger) leads
to an apparent offset between the X-ray centroid and galaxy
distribution. 

\subsection{X-ray surface brightness modeling.}
The two-dimensional (2D) X-ray surface brightness distribution in
ClJ0152-7$-$1357 was modeled in \emph{Sherpa} with an elliptical
$\beta-$model for each of the two subclusters and two groups, and a flat
and a vignetted background \citep[see][for a more detailed description of
the method]{mau04a}. Due to the low surface brightness of the two groups,
their model ellipticities were fixed at zero and their core radii were
fixed at $100\kpc$. This core radius is appropriate for the temperatures of
these groups as measured in \textsection \ref{sect.specmap}
\citep{san03}. The models were fit simultaneously to binned images for each
EPIC camera after being convolved with the appropriate PSF and exposure map. All model parameters except for the
amplitudes were tied between the different detectors. The background levels
were determined from a fit to a local source free region and fixed.

\begin{figure}
\begin{center}
\plotone{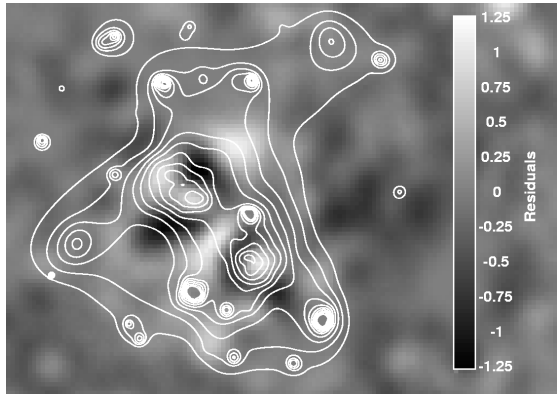}
\caption{\label{fig:resids}Smoothed X-ray image of the residuals from the
best-fitting model to the surface brightness distribution in
ClJ0152-7$-$1357. Smoothed contours of X-ray surface brightness are
overlaid.} 
\end{center}
\end{figure}

\begin{figure*}
\begin{center}
\plottwo{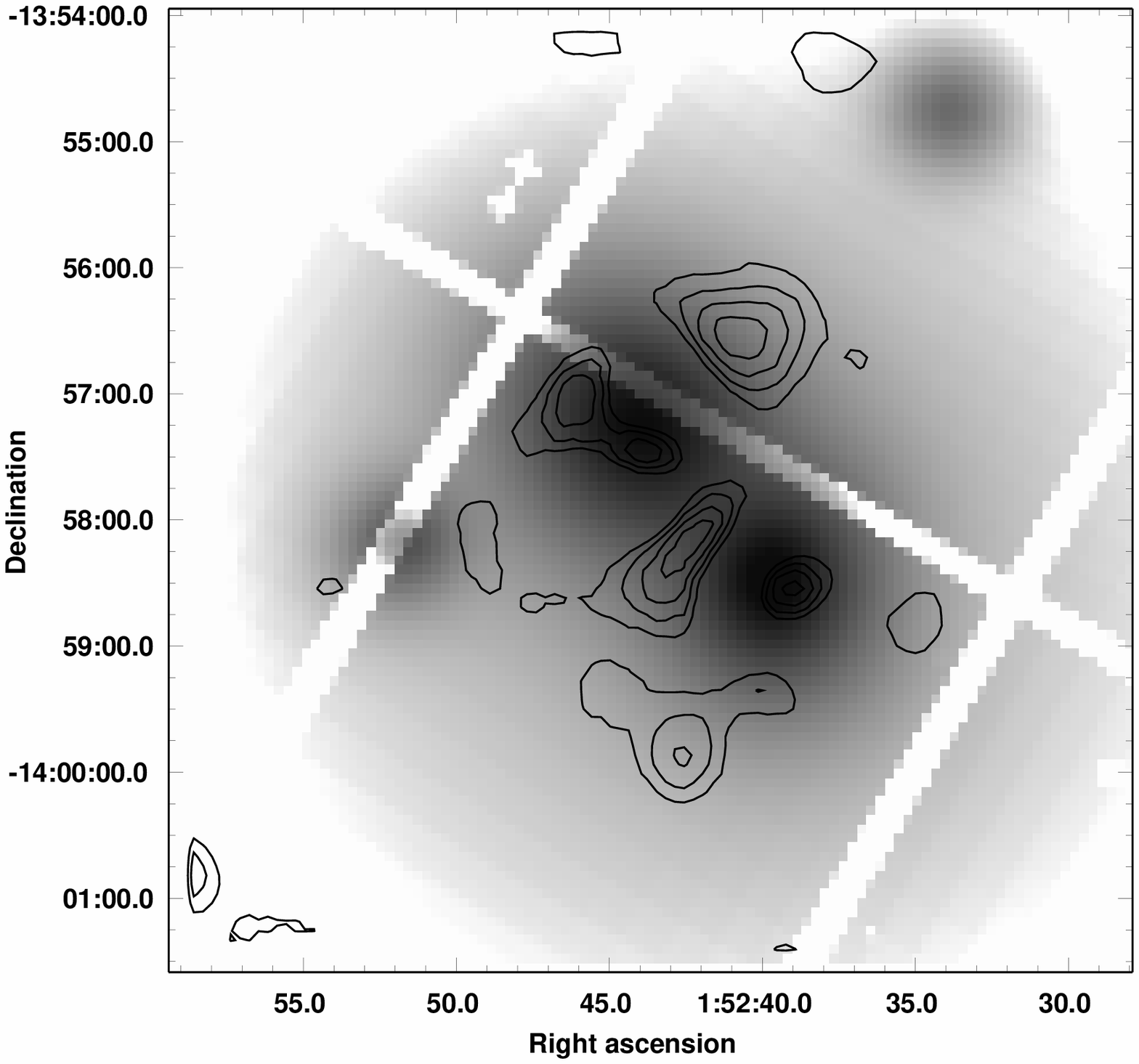}{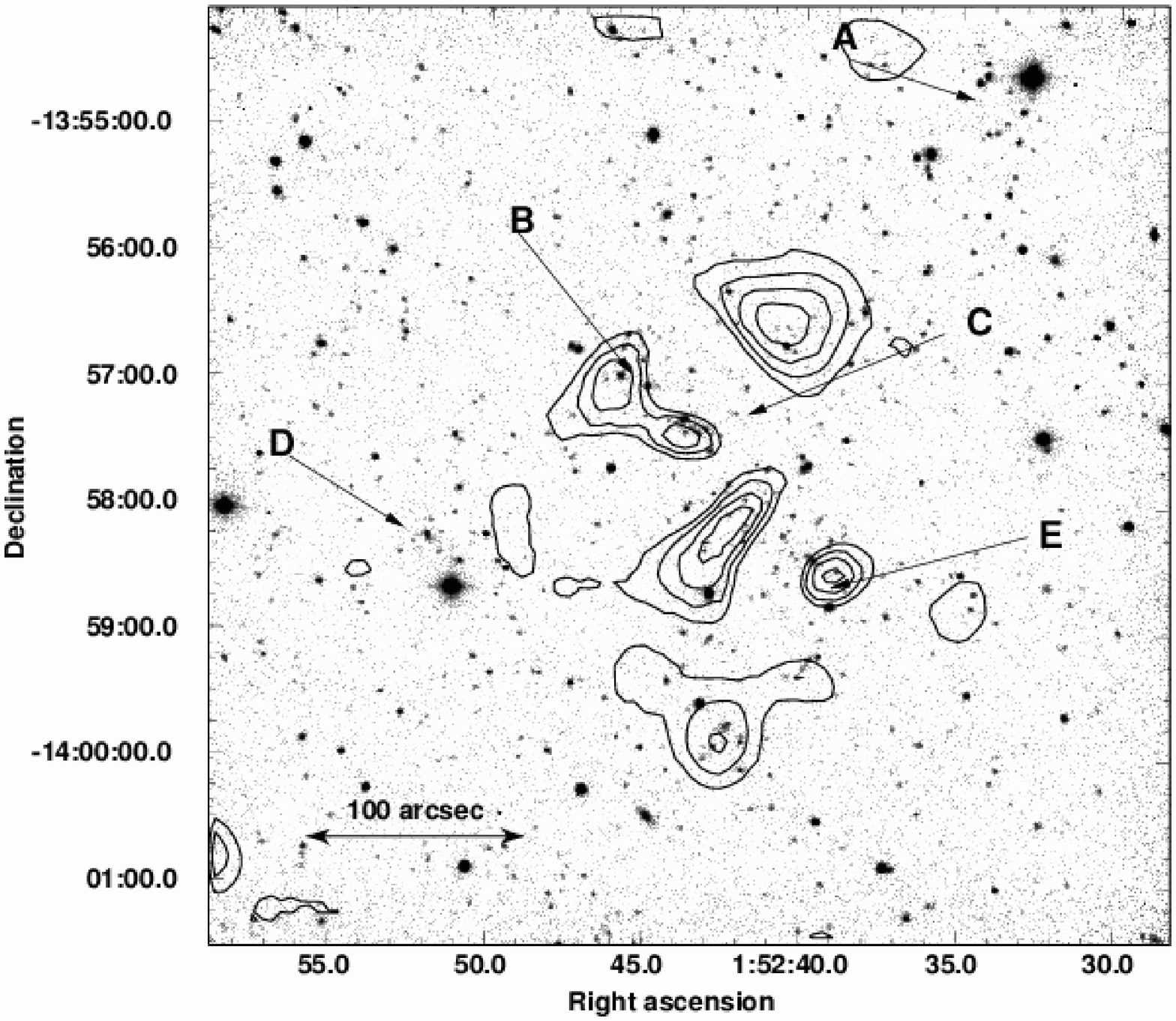}
\caption{\label{fig:residoverlay}{\it Left:} The best-fitting
two-dimensional surface
brightness model of \jjj\ with contours of the positive residuals
overlaid. {\it Right:} The same contours are overlaid on an IRIS2
K band image, with the same labels used in Fig \ref{fig:overlay}.} 
\end{center}
\end{figure*}

The data residuals from the best-fitting models for each camera were
combined and smoothed with a Gaussian of $\sigma=12\arcs$ (3 image
pixels). The resulting image is shown in Fig. \ref{fig:resids}. Point
sources were excluded both during the fitting process and from these
residuals. There are several regions of excess emission above the model
visible in Fig. \ref{fig:resids}. We note that some of these residual
features occur near or across PN CCD gaps. However, they are also present
in the residuals for the MOS detectors alone so are not due to exposure
correction problems around the CCD gaps. 
In Fig. \ref{fig:residoverlay}, contours of the positive surface brightness
residuals are overlaid on the best-fitting surface brightness model, and on
a NIR image. In this Figure, the surface brightness models for each
detector have been summed, after convolution with the PSF and
multiplication by the exposure map. The PN CCD gaps are thus visible in the
surface brightness model. The same labels used in Fig.
\ref{fig:overlay} are included to aid orientation. Positive residuals are
associated with the subcluster cores (C and E), which may indicate the
presence of dense cool cores, and also with region B. The region of excess
emission at $\alpha[2000.0]=01^{\rm h}52^{\rm m}40.75^{\rm s}$
$\delta[2000.0]=-13^{\circ}56\arcm 33.1\arcs$ to the west of region B is
aligned with several faint galaxies, while that at $\alpha[2000.0]=01^{\rm
h}52^{\rm m}42.71^{\rm s}$ $\delta[2000.0]=-13^{\circ}59\arcm 52.5\arcs$ to
the south east of the southern subcluster (E) coincides with several
brighter galaxies. Both features appear to be associated with galaxies
whose photometric redshifts are consistent with that of \jjj\ as measured
by \citet{kod05}, and both appear in the contours of galaxy light in
Fig. \ref{fig.galcont}.  While this is not statistically rigorous, it is
likely that these regions of excess emission are linked to levels of
substructure in the system which are not present in our simple surface
brightness model. Similarly, the negative residuals to either side of the
northern subcluster core in Fig. \ref{fig:resids} are probably due to the
inadequacy of a simple elliptical model to describe the subcluster.

There is excess emission between the the two main
subclusters (C and E). This excess was also detected in the \Chandra\
observation of \jjj\ by \citet{mau03a}. A possible explanation for this
feature proposed by \citet{mau03a} is a region of increased density due to
the compression of the gas in the merger. However, the weak lensing mass
reconstruction of this system by \citet{jee05} shows two mass peaks in this
region, and there are also several cluster galaxies in this region (see
Fig. \ref{fig:residoverlay}). This suggests that the excess emission may be
due to gas associated with mass clumps in this region, which may be
additional subgroups or clusters in this complex system.    
\subsection{Projections of the X-ray surface brightness}\label{sect.proj}
An alternative way of visualising the data is to use projections of the
surface brightness along different axes. In the case of \jjj\ the two
merger axes are natural choices. A rectangular region aligned with each of
the merger axis was defined. The X-ray counts in a combined PN and MOS
$0.3-5\keV$ image within this region were then projected onto its long
axis. The same procedure was applied to combined PN and MOS image of the
best-fit 2D model, after convolution with the PSF and exposure
map. The data projection was adaptively binned to give fractional
uncertainties of $\le10\%$ on each point. 

\begin{figure}
\begin{center}
\plotone{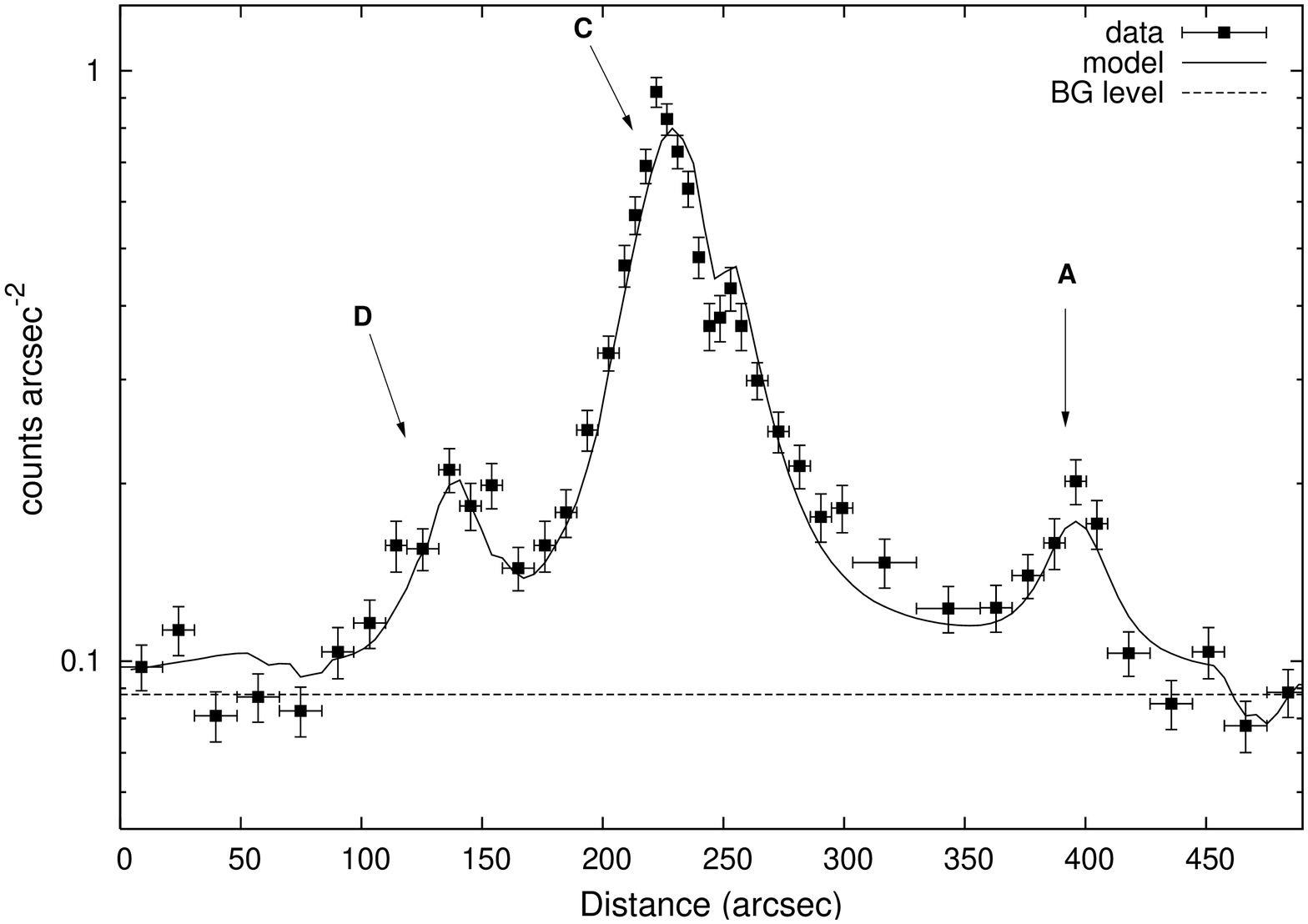}
\caption{\label{fig.SENW}Projection of the surface brightness in a
$75\arcs$ wide region onto the south-east to north-west merger axis. The
points show the projection of the imaging data and the solid line is the
projection of the best-fit 2D model. The dashed line shows the locally
estimated background level.} 
\end{center}
\end{figure}

\begin{figure}
\begin{center}
\plotone{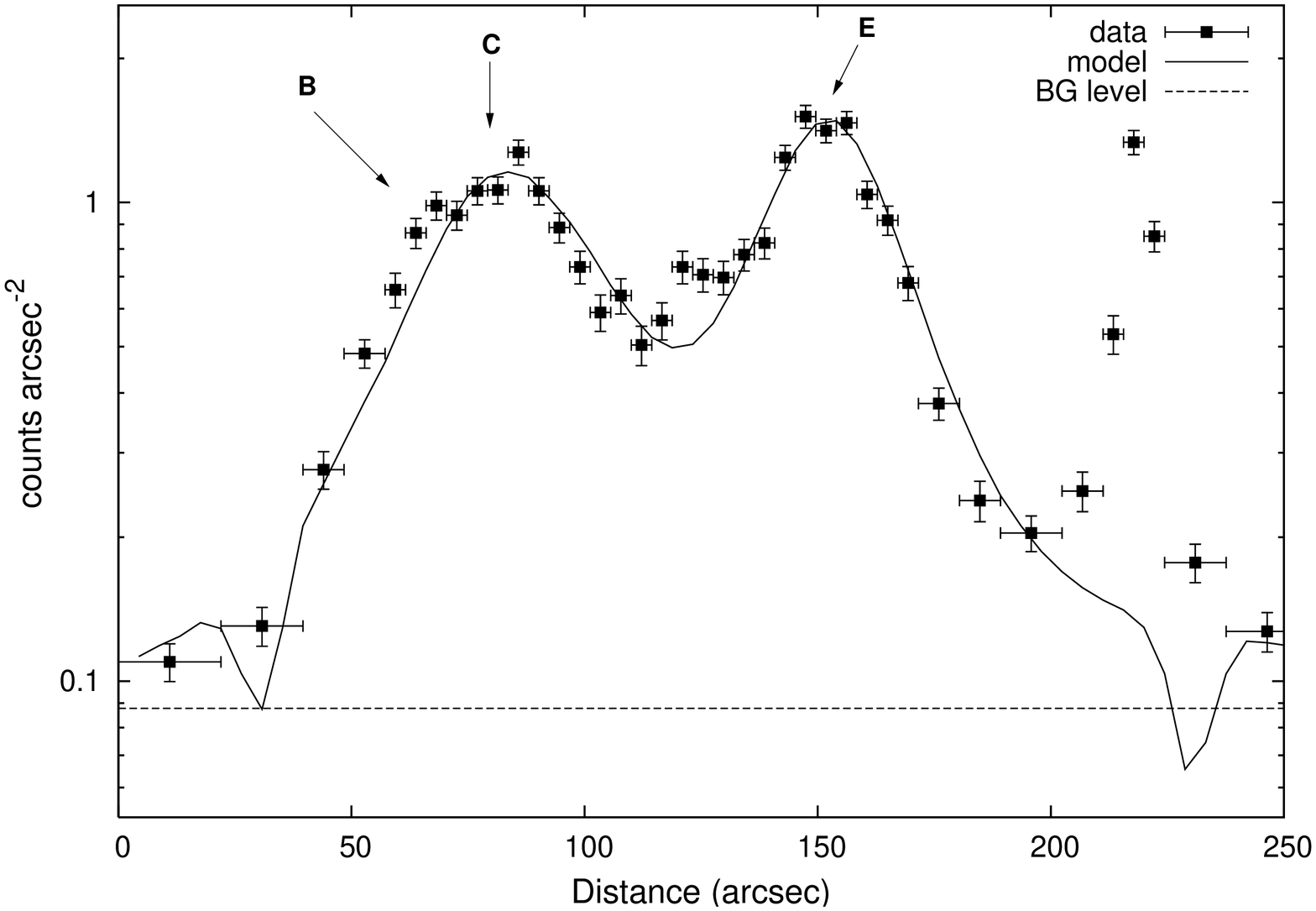}
\caption{\label{fig.NESW}Projection of the surface brightness in a
$50\arcs$ wide region onto the north-east to south-west axis. The
points show the projection of the imaging data and the solid line is the
projection of the best-fit 2D model. The dashed line shows the locally
estimated background level.} 
\end{center}
\end{figure}

Fig. \ref{fig.SENW} shows the projection along the south-east to
north-west merger axis. For this axis, a region of width $75\arcs$ centred
on the two groups was used. Several features are labelled in
Fig. \ref{fig.SENW}, using the same labels as Fig. \ref{fig:overlay} and
Table \ref{tab.labels}. The region between C and A  corresponds to the
apparent filament in Fig. \ref{fig:overlay}. In this region, the model
represents the simple superposition of the group and cluster
emission and the data show no significant departure from this model. The
filamentary appearance of the emission in this region could thus simply be
due to the superposition of the cluster and group emission. However, given the
difficulty of modeling the surface brightness distribution of this system,
we cannot rule out the possibility of filamentary emission in this
region. The point source visible in Fig. \ref{fig:overlay} halfway between
B and A was excluded from the data and model in this projection.
The projection of the north-east to south-west merger axis is shown in
Fig. \ref{fig.NESW}. A rectangular projection region of width $50\arcs$
centred on the two subclusters was used. The substructure around region B
in Fig. \ref{fig:overlay} is visible in this projection. Excess emission
between the subclusters (C and E) is apparent, and a bright point source
which was excluded from the 2D modeling, but not from the data in this
projection is visible at the right end of Fig. \ref{fig.NESW}. The
disagreement between the model and data at the position of the chip gap at
$230\arcs$ is due to the imperfect exposure correction around the chip gap
combined with the bright point source.

\section{Spectral Mapping}\label{sect.specmap}
The high quality X-ray data enable spatially resolved temperature
measurements of the gas in this merger. A method based on that described by
\citet{osu05} was used to produce the temperature map. First, a ``radius
map'' was produced, based on a binned X-ray image. The radius map recorded
the radius of the circular region enclosing $>1000$ background-subtracted,
point-source-excluded photons centered on each pixel in the binned image. A
maximum radius of $60\arcs$ was imposed, and pixels which failed to meet
these criteria were excluded from the radius map. Spectra were then
extracted for each pixel within the appropriate radius, and the three EPIC
spectra were fit simultaneously with an absorbed \MEKAL\ model. During the
fitting process, the model parameters were tied except for the
normalisations which were independent. The metal abundance was fixed at
$0.3\Zsol$ using the abundance tables of \citet{and89}, and the absorbing
column was fixed at the Galactic value \citep{dic90}. All spectra were
grouped to have at least 20 total counts per spectral bin, and were fit in
the $0.3-7\keV$ band using local background spectra extracted from the
background annulus region (Fig. \ref{fig.fov}).

\begin{figure*}
\begin{center}
\plotone{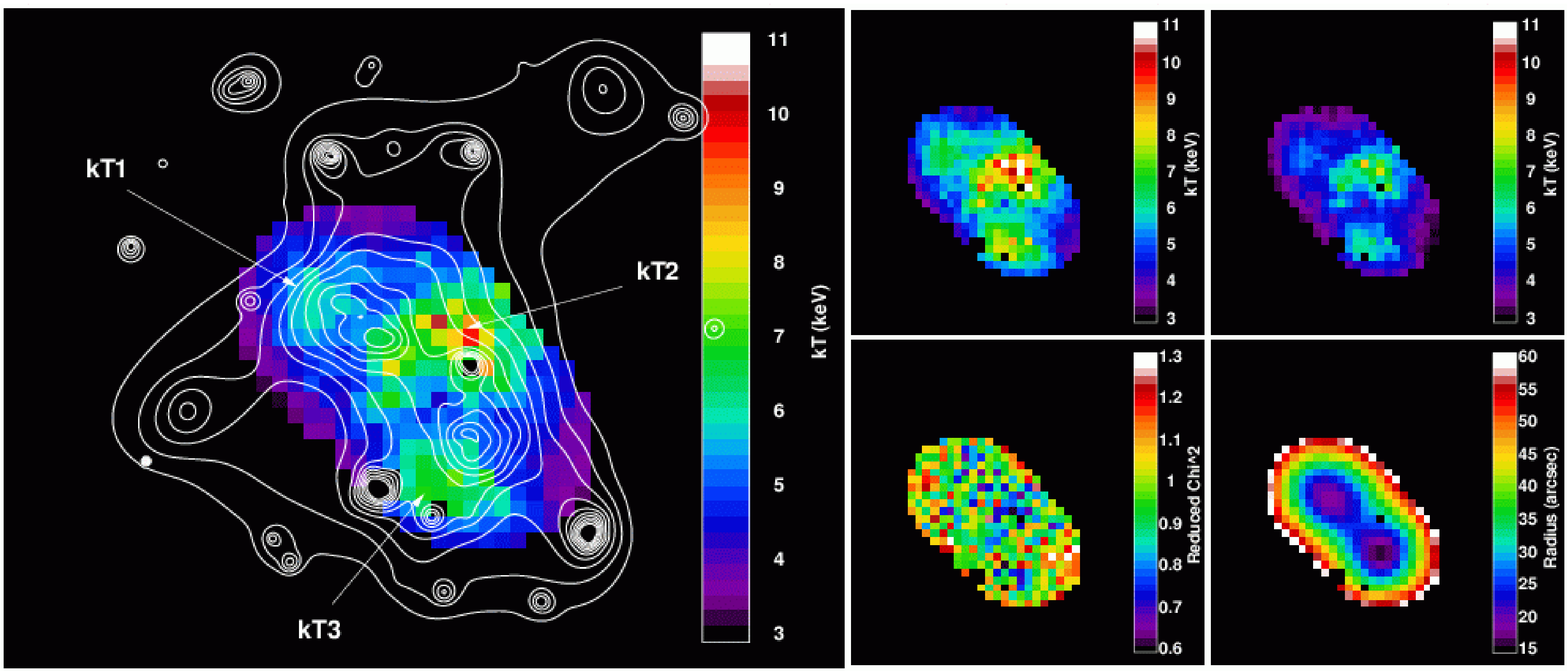}
\caption{\label{fig:ktmap}{\it Left:} Projected temperature map of
ClJ0152-7$-$1357 with contours of X-ray surface brightness overlaid. The
spectral fit parameters of the labelled regions are given in Table
\ref{tab:ktdat}. {\it Right:} The upper panels show the $1\sigma$ upper
(left) and lower (right) limits on the temperature values. The lower panels
show the reduced $\chisq$ of the spectral fits and the radii used for
spectral extraction.} 
\end{center}
\end{figure*}

The resulting projected temperature map is shown in
Fig. \ref{fig:ktmap}. It should be noted that the method used for producing
the temperature map means that adjacent pixels are not independent. The
radius map in Fig. \ref{fig:ktmap} gives the extent of the spectral extraction
region used at each pixel. Broadly speaking, the gas in the system has
temperatures in the range $4-7\keV$ but the spatial distribution of
temperatures is not smooth. There are three apparently hotter regions in
the temperature structure which are labelled in Fig. \ref{fig:ktmap}. The
northern-most (kT1) is aligned with the residual emission at region B (see
Figs. \ref{fig:resids} and \ref{fig:residoverlay}). The hotter regions at
kT2 and kT3 are not associated with any surface brightness features. The
mixing of spectra between adjacent pixels in this spectral mapping method
can make such features appear more significant than they truly are, as they
occur in several adjacent pixels. The temperatures and fit parameters of
the hottest pixel in each of these regions are given in Table
\ref{tab:ktdat}. While region kT1 is a local maximum in temperature, it is
actually no hotter than the average temperature across the cluster of
$5.5\keV$. Regions kT2 and kT3, however, are hotter at the $\sim2\sigma$
level. With the exception of pixels close to the boundaries of the
temperature map, none of the measured temperatures are significantly cooler
than $5.5\keV$.

\begin{figure}
\begin{center}
\plotone{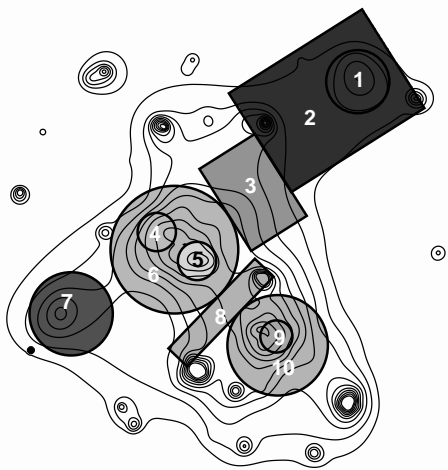}
\caption{\label{fig:geomap}Projected temperature map of
ClJ0152-7$-$1357 with contours of X-ray surface brightness overlaid. Darker
regions are cooler and lighter regions hotter. The fit parameters of the
numbered regions are given in Table \ref{tab:ktdat}.} 
\end{center}
\end{figure}

A simple alternative method of producing a temperature map was also
implemented. In this method geometrical regions were defined by hand using
the surface brightness distribution and residuals as a guide. Spectra were
extracted from these regions (with point sources excluded) and fit as
above. Again, vignetting-corrected background spectra extracted from the
background annulus region were used. The goal in following this method was
to isolate some of the features in the system and obtain spectra that are
free from the mixing effects of the previous method. We were also able to
extract and fit spectra for regions which did not pass the binning
algorithm's criteria. Fig. \ref{fig:geomap} shows the resulting temperature
map and the fit parameters of the numbered regions are given in Table
\ref{tab:ktdat}. The table also gives the net source counts detected with
each EPIC camera. We note that three of the regions (1, 2 and 7) have low
net counts ($150-300$). This means that the grouping of the spectra to 20
total counts per bin and the use of the $\chisq$ statistic may not be
appropriate. In these cases, ungrouped versions of the spectra were also
fit using the C statistic in {\it XSPEC}. In all cases, the temperatures
and uncertainties were in good agreement with the ones quoted in Table
\ref{tab:ktdat}. 

\begin{deluxetable}{ccccccc}
\tablecaption{\label{tab:ktdat}Values of the spectral fit parameters for
the regions shown in Fig. \ref{fig:ktmap} and Fig. \ref{fig:geomap}}
\tablehead{
 &  & \multicolumn{3}{c}{Net Counts} &  &  \\
\cline{3-5}
\colhead{Region} & \colhead{kT (keV)} & PN & MOS1 & MOS2 & \colhead{$\chisq$} & \colhead{$\nu$}} 
\startdata
kT1 & $5.9^{+0.9}_{-0.6}$ & 579 & 240 & 276 & 50.8 & 59 \\
kT2 & $10.3^{+3.0}_{-1.9}$ & 549 & 244 & 245 & 46.0 & 52 \\
kT3 & $7.3^{+1.4}_{-0.9}$ & 705 & 250 & 228 & 44.1 & 59 \\ \hline
1 & $1.5^{+0.2}_{-0.1}$ & 89 & 31 & 38 & 18.5 & 12 \\
2 & $1.4^{+0.3}_{-0.1}$ & 150 & 18 & 28 & 102.8 & 63 \\
3 & $4.4^{+1.0}_{-0.8}$ & 340 & 100 & 123 & 48.4 & 51 \\
4 & $5.6^{+1.1}_{-0.8}$ & 355 & 108 & 135 & 20.6 & 27 \\
5 & $6.4^{+0.8}_{-0.7}$ & 625 & 275 & 233 & 52.0 & 53 \\
6 & $5.4^{+0.4}_{-0.4}$ & 1425 & 620 & 682 & 157.7 & 159 \\
7 & $2.4^{+0.6}_{-0.4}$ & 179 & 76 & 61 & 26.5 & 31 \\
8 & $5.3^{+0.8}_{-0.7}$ & 506 & 161 & 170 & 42.6 & 44 \\
9 & $4.9^{+0.6}_{-0.5}$ & 499 & 188 & 212 & 45.4 & 38 \\
10 & $5.3^{+0.4}_{-0.4}$ & 1590 & 492 & 565 & 142 & 144 
\enddata
\end{deluxetable}

This method indicates that the two main subclusters and their surrounding
regions have temperatures consistent with $5.5\keV$. The eastern group
(region 7) is significantly cooler at $2.4^{+0.6}_{-0.4}\keV$, with a
temperature, appropriate for a group or small cluster. The redshift of the
north-west group (region 1) cannot be determined from the X-ray spectra,
but under the assumption that it is at the cluster redshift it too has a low
temperature consistent with a galaxy group. The emission in region 2,
between group A and the northern subcluster also has a low temperature of
$1.4^{+0.3}_{-0.1}\keV$ assuming a redshift of $0.83$. 

The regions associated with positive surface brightness residuals in the
subcluster cores show no evidence for the cooler gas which would be
expected if the excess emission were due to cool cores. However, the size
of the regions are small ($\sim30\arcs$ diameter) so the blurring effect of
the \XMM\ PSF would make such emission more difficult to detect. The
\Chandra\ observation of this system, with its negligible PSF, also detects
positive surface brightness residuals in the cluster cores, but the data
quality is insufficient to extract spectra from those regions. The
statistical uncertainties on the \XMM\ temperatures along with the
unaccounted for projection effects mean that the presence of cool cores in
either subcluster cannot be ruled out.

Regions 3, 4 and 8 were also chosen based on the surface brightness
residuals. These do not show departures from the subcluster
temperatures. In particular, region 8 between the subclusters is not
significantly hotter.

Finally, the global luminosity of the cluster was measured. Due to the
complex morphology of the system, extrapolating luminosities measured from
spectral fits to the individual components of the system out to a large
radius may lead to inaccuracies. Instead, global spectra were extracted for
each EPIC camera from within a radius of $1.4\Mpc$ \citep[$184\arcs$,
chosen to match the apertures used in][]{mar98a}, after the exclusion of
point sources. The resulting spectra were fit as before using the local
background spectra, and the best fitting temperature was $3.7\pm0.2\keV$
($\chisq/\nu=588/558$). While the emission within this large aperture is
clearly not isothermal, the spectra did not support the modelling of a
second thermal component. The bolometric luminosity (rest frame
$0.01-100\keV$) within this aperture was measured to be
$1.3\pm0.2\times10^{45}\ergps$. The uncertainties on the luminosity
incorporate those on the best-fit temperature and normalisation of the
spectral model.

\section{Discussion}
This deep \XMM\ observation confirms the global temperatures measured by
\Chandra\ for the two main subclusters \citep{mau03a}, and supports the
conclusion that \jjj\ is a massive merger. \citet{mau03a} included just the
two main subclusters in their mass analysis, but the \XMM\ temperatures of
the two groups can be used to estimate their contribution to the total mass
of the system. Using the high-redshift mass-temperature relation of
\citet{mau05a} (based on a sample of relaxed and unrelaxed clusters
including the two subclusters of \jjj) we estimate that an object of
$\sim2\keV$ will have a total virial mass of $\sim1\times10^{14}\Msol$. In
the same way, the approximate temperature of the two subclusters
($5.5\keV$) gives a mass of $\sim5\times10^{14}\Msol$ for each. We thus
concluded that the mass of the system is $\sim1\times10^{15}\Msol$
\citep[in line with][]{mau03a} and that this is dominated by the two main
subclusters.

In the following sections we discuss the important features of \jjj\
detected by \XMM.

\subsection{Substructure in the northern subcluster}
The combination of the galaxy overdensity, weak-lensing mass peak and the
X-ray substructure in the north of the northern subcluster (region B in
Fig. \ref{fig:overlay}) suggest the presence of a third subcluster or group
in the NE-SW merger axis of the \jjj.  One plausible explanation for the
configuration of the galaxy, gas, and dark matter distribution in this
region is that the third subcluster has recently passed through the
northern subcluster, traveling in a northerly direction. The dark matter
and galaxies of the third subcluster were unaffected by the encounter due
to their small (or non-existent) collisional cross sections. Its gas,
meanwhile, was stripped from the potential, with the northern extension of
the X-ray emission and possible second X-ray peak in that region being due
to the surviving dense gas core of the third subcluster. 

The mean redshift of the galaxies associated with this possible third
subcluster is $0.838\pm0.006$ \citep[using the redshifts measured
by][]{dem05}; the same as that of the
northern subcluster itself within the uncertainties. This suggests that any
motion is in the plane of the sky.

\subsection{The properties of group D}
In their dynamical analysis of \jjj, \citet{gir05} find a velocity
dispersion of $\sigma_{\mathrm v}\approx700\kmps$. The X-ray
temperature of the group ($2.4^{+0.6}_{-0.4}\keV$) is slightly low
compared to the velocity dispersion \citep[\egc][]{xue00b}, but it is
consistent with the $\sigma_{\mathrm v}-T$ relation within the
scatter. Using the \MEKAL\ normalisation of the spectral fit in this
region, and assuming that the group is spherical, the gas mass within a
radius of $20\arcs$ was estimated to be $6.0\pm0.6\times10^{11}\Msol$. This
can be compared with the total mass within the same radius from the weak
lensing data \citep{jee05} to give a gas mass fraction of
$\fgas\approx0.03$. At $z=0.83$ the radius of $20\arcs$ corresponds to
$150\kpc$, so this gas fraction is measured only in the central region of
group D. The gas fraction found is consistent with that measured in the
central regions of local systems of a similar temperature
\citep{san03}.

\subsection{Group A and the possible filament}\label{sect.filament}
Another interesting feature of the \XMM\ data is the possibility that
galaxy group A is falling into the cluster along a merger
axis which runs through the northern subcluster (C) and group D. The K-band light
distribution shows similar structure to the  X-ray morphology in this region
and the galaxies associated with the group are faint. However it is not
known if the group is at the same redshift as the cluster. 

The gas density in the region of emission between group A and the northern
subcluster was estimated using the \MEKAL\ normalisation from the spectral
fit to region 2 (Fig. \ref{fig:geomap}). In order to simplify the
computation of the volume of region 2, the \MEKAL\ normalisation was first
scaled up to account for the area excluded in region 1. The volume of
region 2 was then calculated assuming it is a cylinder with rotational
symmetry about its long axis. With the simplifying assumption that the
density is constant throughout that volume, the Hydrogen number density in
this region is $4.0\pm0.6\times10^{-4}\pcc$. These results should be
treated with some caution as the emission is very low surface brightness,
and the issue of background subtraction is critical. In particular, while
the results above were based on spectral fits with a local background, we were
unable to obtain acceptable spectral fits with a blank sky
background. However, the uncertainties in accounting for the differences in
soft Galactic X-ray emission in the source and blank sky datasets are
particularly important in the study of a region of low surface brightness,
cool gas such as this. As discussed in \textsection \ref{sect.data}, the
use of a local background should to be the most reliable method in this
instance.

It is interesting to compare this possible filament with the apparently
isolated filament detected by \citet{sch00} with \ROSAT.  The surface
brightness in region 2 is $1.3\pm0.4\times10^{-15}\flux$
arcmin$^{-2}$. This is an order of magnitude brighter than the
\citet{sch00} filament. While that filament has no apparent connection to
any massive clusters, it is likely that the gas densities would be higher
in regions of filaments that are feeding onto a massive cluster of galaxies
like that in \jjj. However, as demonstrated in \textsection
\ref{sect.proj}, the apparent filament may simply be due to the superposition of
the emission from the group and cluster.

In order to estimate of the extent of the northern subcluster, a mass
profile was derived assuming that the gas follows an isothermal
$\beta$-profile in hydrostatic equilibrium with the cluster
potential. While these are clearly gross simplifications for this complex
system, they are sufficient for the purposes of estimating an approximate
virial radius for the northern subcluster. The gas temperature of region 6
in Fig. \ref{fig:geomap} along with the $\beta$-profile parameters from the
2D surface brightness fitting ($r_c=40\pm3\arcs$, $\beta=0.86\pm0.06$) were
then used to estimate a virial radius of $\rt\approx1.3\Mpc$. Here, we
define the virial radius as the radius enclosing a mean density that is a
factor of 200 times the critical density at $z=0.83$. Numerical simulations
show that this overdensity radius approximately separates the virialised
part of clusters from the infalling material surrounding them \citep{nav95}. The
estimated virial radius of the northern subcluster falls
approximately halfway along region 2 in Fig. \ref{fig:geomap}. Disregarding
projection effects, the southern subcluster and group D all fall
within the virial radius of the northern subcluster. If the group
A is associated with the system, it is most likely at an earlier merger
stage, lying outside the estimated virial radius.

\section{conclusions}
\jjj\ is a fascinating system, comprising two main subclusters in an
apparently early stage of merging and possibly two infalling groups. There
also is evidence for late stage merger activity in the northern
subcluster. The formation of a massive cluster is occurring along two main
merger axes which appear perpendicular in the plane of the sky. This
unique system provides a dramatic example of cluster formation via mergers
on different scales and at different stages, providing further compelling
observational support for the hierarchical assembly model of cluster formation.  

\section{Acknowledgments}
We thank Christine Jones for useful discussions of this work. BJM is
supported by NASA through Chandra Postdoctoral Fellowship Award Number
PF4-50034 issued by the Chandra X-ray Observatory Center, which is operated
by the Smithsonian Astrophysical Observatory for and on behalf of NASA
under contract NAS8-03060. SCE acknowledges PPARC support.

\bibliographystyle{apj}
\bibliography{clusters}

\begin{thebibliography}{38}
\expandafter\ifx\csname natexlab\endcsname\relax\def\natexlab#1{#1}\fi

\bibitem[{{Anders} \& {Grevesse}(1989)}]{and89}
{Anders}, E. \& {Grevesse}, N. 1989, \gca, 53, 197

\bibitem[{{Arnaud} {et~al.}(2002){Arnaud}, {Majerowicz}, {Lumb}, {Neumann},
  {Aghanim}, {Blanchard}, {Boer}, {Burke}, {Collins}, {Giard}, {Nevalainen},
  {Nichol}, {Romer}, \& {Sadat}}]{arn02b}
{Arnaud}, M., {Majerowicz}, S., {Lumb}, D., {Neumann}, D.~M., {Aghanim}, N.,
  {Blanchard}, A., {Boer}, M., {Burke}, D.~J., {Collins}, C.~A., {Giard}, M.,
  {Nevalainen}, J., {Nichol}, R.~C., {Romer}, A.~K., \& {Sadat}, R. 2002, \aap,
  390, 27

\bibitem[{{Arnaud} {et~al.}(2000){Arnaud}, {Maurogordato}, {Slezak}, \&
  {Rho}}]{arn00}
{Arnaud}, M., {Maurogordato}, S., {Slezak}, E., \& {Rho}, J. 2000, \aap, 355,
  461

\bibitem[{{Colless} {et~al.}(2001){Colless}, {Dalton}, {Maddox}, {Sutherland},
  {Norberg}, {Cole}, {Bland-Hawthorn}, {Bridges}, {Cannon}, {Collins}, {Couch},
  {Cross}, {Deeley}, {De Propris}, {Driver}, {Efstathiou}, {Ellis}, {Frenk},
  {Glazebrook}, {Jackson}, {Lahav}, {Lewis}, {Lumsden}, {Madgwick}, {Peacock},
  {Peterson}, {Price}, {Seaborne}, \& {Taylor}}]{col01}
{Colless}, M., {Dalton}, G., {Maddox}, S., {Sutherland}, W., {Norberg}, P.,
  {Cole}, S., {Bland-Hawthorn}, J., {Bridges}, T., {Cannon}, R., {Collins}, C.,
  {Couch}, W., {Cross}, N., {Deeley}, K., {De Propris}, R., {Driver}, S.~P.,
  {Efstathiou}, G., {Ellis}, R.~S., {Frenk}, C.~S., {Glazebrook}, K.,
  {Jackson}, C., {Lahav}, O., {Lewis}, I., {Lumsden}, S., {Madgwick}, D.,
  {Peacock}, J.~A., {Peterson}, B.~A., {Price}, I., {Seaborne}, M., \&
  {Taylor}, K. 2001, \mnras, 328, 1039

\bibitem[{Della~Ceca {et~al.}(2000)Della~Ceca, Scaramella, Gioia, Rosati, \&
  Squires}]{del00}
Della~Ceca, R., Scaramella, R., Gioia, I.~M., Rosati, F., \& Squires, G. 2000,
  \aap, 353, 498

\bibitem[{{Demarco} {et~al.}(2005){Demarco}, {Rosati}, {Lidman}, {Homeier},
  {Scannapieco}, {Ben{\'{\i}}tez}, {Mainieri}, {Nonino}, {Girardi}, {Stanford},
  {Tozzi}, {Borgani}, {Silk}, {Squires}, \& {Broadhurst}}]{dem05}
{Demarco}, R., {Rosati}, P., {Lidman}, C., {Homeier}, N.~L., {Scannapieco}, E.,
  {Ben{\'{\i}}tez}, N., {Mainieri}, V., {Nonino}, M., {Girardi}, M.,
  {Stanford}, S.~A., {Tozzi}, P., {Borgani}, S., {Silk}, J., {Squires}, G., \&
  {Broadhurst}, T.~J. 2005, \aap, 432, 381

\bibitem[{Dickey \& Lockman(1990)}]{dic90}
Dickey, J.~M. \& Lockman, F.~J. 1990, \araa, 28, 215

\bibitem[{{Durret} {et~al.}(2005){Durret}, {Lima Neto}, \& {Forman}}]{dur05}
{Durret}, F., {Lima Neto}, G.~B., \& {Forman}, W. 2005, \aap, 432, 809

\bibitem[{{Durret} {et~al.}(2003){Durret}, {Lima Neto}, {Forman}, \&
  {Churazov}}]{dur03}
{Durret}, F., {Lima Neto}, G.~B., {Forman}, W., \& {Churazov}, E. 2003, \aap,
  403, L29, lSS

\bibitem[{{Ebeling} {et~al.}(2004){Ebeling}, {Barrett}, \& {Donovan}}]{ebe05a}
{Ebeling}, H., {Barrett}, E., \& {Donovan}, D. 2004, \apjl, 609, L49

\bibitem[{Ebeling {et~al.}(2000)Ebeling, Jones, Perlman, Scharf, Horner,
  Wegner, Malkan, Fairley, \& Mullis}]{ebe00a}
Ebeling, H., Jones, L.~R., Perlman, E., Scharf, C., Horner, D., Wegner, G.,
  Malkan, M., Fairley, B., \& Mullis, C.~R. 2000, \apj, 534, 133

\bibitem[{{Ebeling} {et~al.}(2005){Ebeling}, {White}, \& {Rangarajan}}]{ebe05b}
{Ebeling}, H., {White}, D., \& {Rangarajan}, F. 2005, \mnras

\bibitem[{{Ferrari} {et~al.}(2005){Ferrari}, {Arnaud}, {Ettori},
  {Maurogordato}, \& {Rho}}]{fer05}
{Ferrari}, C., {Arnaud}, M., {Ettori}, S., {Maurogordato}, S., \& {Rho}, J.
  2005, ArXiv Astrophysics e-prints

\bibitem[{{Gavazzi} {et~al.}(1996){Gavazzi}, {Pierini}, \& {Boselli}}]{gav96}
{Gavazzi}, G., {Pierini}, D., \& {Boselli}, A. 1996, \aap, 312, 397

\bibitem[{{Girardi} {et~al.}(2005){Girardi}, {Demarco}, {Rosati}, \&
  {Borgani}}]{gir05}
{Girardi}, M., {Demarco}, R., {Rosati}, P., \& {Borgani}, S. 2005, \aap, 442,
  29

\bibitem[{{Huo} {et~al.}(2004){Huo}, {Xue}, {Xu}, {Squires}, \&
  {Rosati}}]{huo04}
{Huo}, Z., {Xue}, S., {Xu}, H., {Squires}, G., \& {Rosati}, P. 2004, \aj, 127,
  1263

\bibitem[{{Jee} {et~al.}(2005){Jee}, {White}, {Ben{\'{\i}}tez}, {Ford},
  {Blakeslee}, {Rosati}, {Demarco}, \& {Illingworth}}]{jee05}
{Jee}, M.~J., {White}, R.~L., {Ben{\'{\i}}tez}, N., {Ford}, H.~C., {Blakeslee},
  J.~P., {Rosati}, P., {Demarco}, R., \& {Illingworth}, G.~D. 2005, \apj, 618,
  46

\bibitem[{{Jeltema} {et~al.}(2005){Jeltema}, {Canizares}, {Bautz}, \&
  {Buote}}]{jel05}
{Jeltema}, T.~E., {Canizares}, C.~R., {Bautz}, M.~W., \& {Buote}, D.~A. 2005,
  \apj, 624, 606

\bibitem[{{Jenkins} {et~al.}(1998){Jenkins}, {Frenk}, {Pearce}, {Thomas},
  {Colberg}, {White}, {Couchman}, {Peacock}, {Efstathiou}, \& {Nelson}}]{jen98}
{Jenkins}, A., {Frenk}, C.~S., {Pearce}, F.~R., {Thomas}, P.~A., {Colberg},
  J.~M., {White}, S.~D.~M., {Couchman}, H.~M.~P., {Peacock}, J.~A.,
  {Efstathiou}, G., \& {Nelson}, A.~H. 1998, \apj, 499, 20

\bibitem[{Joy {et~al.}(2001)Joy, LaRoque, Grego, Carlstrom, Dawson, Ebeling,
  Holzapfel, Nagai, \& Reese}]{joy01}
Joy, M., LaRoque, S., Grego, L., Carlstrom, J.~E., Dawson, K., Ebeling, H.,
  Holzapfel, W.~L., Nagai, D., \& Reese, E.~D. 2001, \apj, 551, L1

\bibitem[{{Kodama} {et~al.}(2005){Kodama}, {Tanaka}, {Tamura}, {Yahagi},
  {Nagashima}, {Tanaka}, {Arimoto}, {Futamase}, {Iye}, {Karasawa}, {Kashikawa},
  {Kawasaki}, {Kitayama}, {Matsuhara}, {Nakata}, {Ohashi}, {Ohta}, {Okamoto},
  {Okamura}, {Shimasaku}, {Suto}, {Tamura}, {Umetsu}, \& {Yamada}}]{kod05}
{Kodama}, T., {Tanaka}, M., {Tamura}, T., {Yahagi}, H., {Nagashima}, M.,
  {Tanaka}, I., {Arimoto}, N., {Futamase}, T., {Iye}, M., {Karasawa}, Y.,
  {Kashikawa}, N., {Kawasaki}, W., {Kitayama}, T., {Matsuhara}, H., {Nakata},
  F., {Ohashi}, T., {Ohta}, K., {Okamoto}, T., {Okamura}, S., {Shimasaku}, K.,
  {Suto}, Y., {Tamura}, N., {Umetsu}, K., \& {Yamada}, T. 2005, \pasj, 57, 309

\bibitem[{Markevitch(1998)}]{mar98a}
Markevitch, M. 1998, \apj, 504, 27

\bibitem[{Markevitch {et~al.}(2002)Markevitch, Gonzalez, Vikhlinin, Murray,
  Forman, Jones, \& Tucker}]{mar02}
Markevitch, M., Gonzalez, L., Vikhlinin, A., Murray, S., Forman, W., Jones, C.,
  \& Tucker, W. 2002, \apj, 567, L27

\bibitem[{{Markevitch} {et~al.}(2000){Markevitch}, {Ponman}, {Nulsen}, {Bautz},
  {Burke}, {David}, {Davis}, {Donnelly}, {Forman}, \& {Jones}}]{mar00c}
{Markevitch}, M., {Ponman}, T.~J., {Nulsen}, P.~E.~J., {Bautz}, M.~W., {Burke},
  D.~J., {David}, L.~P., {Davis}, D., {Donnelly}, R.~H., {Forman}, W.~R., \&
  {Jones}, C. 2000, \apj, 541, 542

\bibitem[{Maughan {et~al.}(2003)Maughan, Jones, Ebeling, Perlman, Rosati, Frye,
  \& Mullis}]{mau03a}
Maughan, B.~J., Jones, L.~R., Ebeling, H., Perlman, E., Rosati, P., Frye, C.,
  \& Mullis, C.~R. 2003, \apj, 587, 589

\bibitem[{{Maughan} {et~al.}(2004){Maughan}, {Jones}, {Ebeling}, \&
  {Scharf}}]{mau04a}
{Maughan}, B.~J., {Jones}, L.~R., {Ebeling}, H., \& {Scharf}, C. 2004, \mnras,
  351, 1193

\bibitem[{{Maughan} {et~al.}(2005){Maughan}, {Jones}, {Ebeling}, \&
  {Scharf}}]{mau05a}
---. 2005, \mnras, in press, (astro

\bibitem[{Navarro {et~al.}(1995)Navarro, Frenk, \& White}]{nav95}
Navarro, J.~F., Frenk, C.~S., \& White, S. D.~M. 1995, \mnras, 275, 720

\bibitem[{{Neumann} {et~al.}(2003){Neumann}, {Lumb}, {Pratt}, \&
  {Briel}}]{neu03}
{Neumann}, D.~M., {Lumb}, D.~H., {Pratt}, G.~W., \& {Briel}, U.~G. 2003, \aap,
  400, 811

\bibitem[{{O'Sullivan} {et~al.}(2005){O'Sullivan}, {Vrtilek}, {Kempner},
  {David}, \& {Houck}}]{osu05}
{O'Sullivan}, E., {Vrtilek}, J.~M., {Kempner}, J.~C., {David}, L.~P., \&
  {Houck}, J.~C. 2005, \mnras, 357, 1134

\bibitem[{{Read} \& {Ponman}(2003)}]{rea03}
{Read}, A.~M. \& {Ponman}, T.~J. 2003, \aap, 409, 395

\bibitem[{Romer {et~al.}(2000)Romer, Nichol, Holden, Ulmer, Pildis, Merrelli,
  Adami, Burke, Collins, Metevier, Kron, \& Commons}]{rom00}
Romer, A.~K., Nichol, R.~C., Holden, B.~P., Ulmer, M.~P., Pildis, R.~A.,
  Merrelli, A.~J., Adami, C., Burke, D.~J., Collins, C.~A., Metevier, A.~J.,
  Kron, R.~G., \& Commons, K. 2000, \apjs, 126, 209

\bibitem[{Rosati {et~al.}(1998)Rosati, Della~Ceca, Norman, \& Giaconni}]{ros98}
Rosati, P., Della~Ceca, R., Norman, C., \& Giaconni, R. 1998, \apj, 492, L21

\bibitem[{{Sanderson} {et~al.}(2003){Sanderson}, {Ponman}, {Finoguenov},
  {Lloyd-Davies}, \& {Markevitch}}]{san03}
{Sanderson}, A.~J.~R., {Ponman}, T.~J., {Finoguenov}, A., {Lloyd-Davies},
  E.~J., \& {Markevitch}, M. 2003, \mnras, 340, 989

\bibitem[{{Scharf} {et~al.}(2000){Scharf}, {Donahue}, {Voit}, {Rosati}, \&
  {Postman}}]{sch00}
{Scharf}, C., {Donahue}, M., {Voit}, G.~M., {Rosati}, P., \& {Postman}, M.
  2000, \apjl, 528, L73

\bibitem[{Scharf {et~al.}(1997)Scharf, Jones, Ebeling, Perlman, Malkan, \&
  Wegner}]{sch97}
Scharf, C., Jones, L.~R., Ebeling, H., Perlman, E., Malkan, M., \& Wegner, G.
  1997, \apj, 477, 79

\bibitem[{{Vikhlinin} {et~al.}(2001){Vikhlinin}, {Markevitch}, \&
  {Murray}}]{vik01}
{Vikhlinin}, A., {Markevitch}, M., \& {Murray}, S.~S. 2001, \apj, 551, 160

\bibitem[{{Xue} \& {Wu}(2000)}]{xue00b}
{Xue}, Y. \& {Wu}, X. 2000, \apj, 538, 65

\end{thebibliography}

\end{document}